\documentclass[11pt]{JHEP3}

\JHEPspecialurl{http://jhep.sissa.it/JOURNAL/JHEP3.tar.gz}

\usepackage{epsfig,multicol}
\usepackage{amsmath}

\newcommand\fverb{\setbox\fverbbox=\hbox\bgroup\verb}
\newcommand\fverbdo{\egroup\medskip\noindent%
			\fbox{\unhbox\fverbbox}\ }
\newcommand\fverbit{\egroup\item[\fbox{\unhbox\fverbbox}]}
\newbox\fverbbox

\newcommand{\pslash}{p\kern-1ex /}
\newcommand{\qslash}{q\kern-1ex /}
\newcommand{\lslash}{l\kern-1ex /}
\newcommand{\sslash}{s\kern-1ex /}
\newcommand{\kaslash}{k_a\kern-2ex /}
\newcommand{\kbslash}{k_b\kern-2ex /}
\newcommand{\Dslash}{\mathcal{D}\kern-1.5ex /}

\newcommand{\beqa}{\begin{eqnarray}}
\newcommand{\eeqa}{\end{eqnarray}}

\newcommand{\ba}{\begin{eqnarray}}
\newcommand{\ea}{\end{eqnarray}}
\newcommand{\be}{\begin{equation}}

\title{Operator product expansion and the short distance
behavior of 3-flavor baryon potentials}
\author{Sinya Aoki\\
       Graduate School of Pure and Applied Sciences, University of Tsukuba, Tsukuba, Ibaraki 305-8571,    Japan, and\\
       Center for Computational Sciences, University of Tsukuba, Tsukuba, Ibaraki 305-8577,    Japan\\
        E-mail: \email{saoki@het.ph.tsukuba.ac.jp}}

\author{Janos Balog\\
        Research Institute for Particle and Nuclear Physics, 
1525 Budapest 114, Pf. 49, Hungary\\
        E-mail: \email{balog@rmki.kfki.hu}}
 \author{Peter Weisz\\
        Max-Planck-Institut f\"ur Physik, F\"ohringer Ring 6, D-80805 M\"unchen, Germany\\
        E-mail: \email{pew@mpp.mpg.de}}       

\received{\today} 		
\accepted{\today}	
\preprint{MPP-2010-59, UTHEP-608}

\abstract{
The short distance behavior of baryon-baryon potentials defined through
Nambu-Bethe-Salpeter wave functions is investigated using the operator
product expansion.
In a previous analysis of the nucleon-nucleon case, 
corresponding to the SU(3) channels $27_s$ and $\overline{10}_a$,
we argued that the potentials have a repulsive core.
A new feature occurs for the case of baryons made up of three flavors:
manifestly asymptotically attractive potentials appear in the singlet 
and octet channels. Attraction in the singlet channel was first indicated
by quark model considerations, and recently been found in numerical lattice
simulations. The latter have however not yet revealed asymptotic attraction 
in the octet channels; we give a speculative explanation for this
apparent discrepancy.
}

\keywords{ Repulsive core, operator product expansion, 3-flavor baryon potential, 
attraction in the singlet channel}

\begin{document}

\section{Introduction}

In a recent paper \cite{IAH} a proposal has been made 
to study nucleon--nucleon (NN) potentials from a first principle 
QCD approach. In this field theoretic framework, potentials are 
obtained through the Schr\"odinger operator applied to
Nambu--Bethe--Salpeter (NBS) wave functions: 
\begin{eqnarray}
V(\vec x) = E +\frac{1}{2\mu}\frac{\nabla^2 
\varphi_E(\vec x)}{\varphi_E(\vec x)}.
\label{pot}
\end{eqnarray}
Here $\mu$ is the two nucleon reduced mass and the NBS wave function is 
defined by
\beqa
\varphi_E(\vec x) &=& \langle 0 \vert N(\vec x/2, 0)
N(-\vec x/2,0) 
\vert 2{\rm N}, E\rangle\,,
\label{wf}
\eeqa
where $\vert 2{\rm N}, E\rangle$ is a QCD eigenstate with energy $E$ 
(suppressing here other quantum numbers),
and $N$ is a nucleon interpolating operator made of 3 quarks.
Such wave functions have been measured through numerical simulations
of the lattice regularized theory \cite{IAH,AHI1,AHI2,IAH2}. 
Although many conceptual questions remain to be resolved
(the potential defined by (\ref{pot}) is for example energy dependent),
the corresponding potentials 
indeed qualitatively resemble phenomenological NN potentials which 
are widely used in nuclear physics. The force at medium to long distance 
($r\ge 2$ fm) is shown to be attractive. This feature has long well been 
understood in terms of pion and other heavier meson exchanges.
At short distance, a characteristic repulsive core is reproduced by the 
lattice QCD simulation \cite{IAH}. 
In \cite{abw3} we performed an operator product expansion (OPE) analysis
of NN NBS wave functions in QCD to theoretically 
better understand the repulsive core of the NN potential
(at least that of the measured NBS potential).  
Thanks to the property of asymptotic freedom of QCD
the form of leading short distance behavior of the coefficent functions can
be computed using perturbation theory. To set the stage we first give
a very short summary of the results of~\cite{abw3}. 

The behavior of the wave functions $\varphi_E(\vec x)$ 
at short distances ($r=\vert \vec x\vert\to0$) 
is encoded in the operator product expansion (OPE) of 
the two nucleon operators: 
\begin{equation}
N(\vec x/2,0)\,N(-\vec x/2,0)\approx\sum_k
D_k(\vec x)\,{\cal O}_k(\vec 0,0),
\end{equation}
where $\{ {\cal O}_k \}$ is a set of local color singlet 6-quark operators 
with two-nucleon quantum numbers. Asymptotically the $\vec x$-dependence 
and energy dependence of the wave function is factorized into
\begin{equation}
\varphi_E(\vec x)\approx\sum_k D_k(\vec x)
\langle0\vert{\cal O}_k(\vec 0,0)\vert 2{\rm N},E\rangle\,.
\end{equation}
Standard renormalization group (RG) analysis  gives \cite{abw3} the leading 
short distance behavior of the OPE coefficient function as
\begin{equation}
D_k(\vec x)\approx \left(\ln\frac{r_0}{r}\right)^{\nu_k}\,d_k\,,
\end{equation}
where $\nu_k$ is related to the 1-loop coefficient of the anomalous dimension
of the operator ${\cal O}_k$, $d_k$ is the tree-level 
contribution\footnote{In principle, the leading term might come from an 
operator that has vanishing tree level coefficient $d_k$ and enters the 
two-nucleon OPE at higher order in perturbation theory, provided that its
anomalous dimension is large enough to compensate. In \cite{abw3} we calculated
the anomalous dimensions of all relevant 6-quark operators and showed that no
such operators exist.}
of $D_k(\vec 0)$ and finally $r_0$ is some typical non-perturbative
QCD scale (which we take here to be 1 fm). Clearly the operator with largest
RG power $\nu_k$ dominates the wave function (\ref{wf}) at short distances.
We will denote the leading (largest) power by $\nu_1$ and the subleading
one (second largest) by $\nu_2$.

If neither $\nu_1$ nor $d_1$ vanishes, this leads to the leading
asymptotics of the s-wave potential of the form
\begin{equation}
V(r)\approx-\frac{\nu_1}{r^2\left(\ln\frac{r_0}{r}\right)}\,,
\end{equation}
which is attractive for $\nu_1>0$ and repulsive for $\nu_1<0$.
Note that the above asymptotic form is energy independent.

If $\nu_1=0$, the situation is more complicated. In this
case the relative sign of the ratio $R$ between the leading contribution and  the subleading 
contribution, corresponding to an operator ${\cal O}_2$ with $\nu_2<0$,
is important.  If $R$ is positive, the potential is repulsive, while it is
attractive for negative $R$. The leading asymptotics of the potential 
is energy dependent in this case, since it is proportional to $R$ which
depends on the energy and it may even change sign as function of $E$. 

In the nucleon potential problem the above degenerate case ($\nu_1=0)$
occurs. In \cite{abw3} we have argued that the relative coefficient $R$
is positive and the short distance limit of the nucleon potential is 
repulsive in both possible isospin-spin s-wave channels ($IJ=01$ 
and $IJ=10$).         

To actually compute the renormalization group power $\nu_k$ of an 
operator, we first have to
compute the 1-loop coefficient of its anomalous dimension defined by 
\cite{abw3}:
\begin{equation}
\gamma_k(g)=\gamma^{(1)}_k\,g^2+{\rm O}(g^4)\,.
\end{equation}
Here we assume that the corresponding operator ${\cal O}_k$ is renormalized
multiplicatively. It is always possible to choose an operator basis where this
is the case, at least up to 1-loop level. Having computed the coefficient
$\gamma^{(1)}_k$ the RG power $\nu_k$ is given by 
\begin{equation}
\nu_k=\frac{\gamma_k^{(1)}-2\gamma_N^{(1)}}{2\beta_0} =\frac{48\pi^2\,\gamma^{(1)}_k-24}{66-4N_f}\,,
\label{betak}
\end{equation}
where $\beta_0$ is the 1-loop beta function coefficient of the $N_f$ flavor theory
and $\gamma_N^{(1)}=12/48\pi^2$ is the 1-loop coefficient of the anomalous dimension of the nucleon operator.  For simplicity we count the anomalous dimension 
coefficients in units of $1/48\pi^2$ and for example we shall call an 
operator ${\cal O}_k$ of anomalous dimension 36 if 
$\gamma^{(1)}_k=\frac{3}{4\pi^2}=\frac{36}{48\pi^2}$.

In this paper we extend the analysis of ref. \cite{abw3} to the case of three
flavors. Instead of nucleons, we shall consider the baryon octet and study the
short distance behavior of the NBS wave function
\begin{equation}
\varphi^{ij}_{\alpha\beta}(\vec x)=
\langle0\vert B^i_\alpha(\vec x/2,0)\,B^j_\beta(-\vec x/2,0)\vert
2{\rm B},E\rangle\,,
\end{equation}
where $i,j$ are SU$(3)$ flavor octet indices, $\alpha,\beta$ spin indices,
$B^i_\alpha$ baryon field operators and $\vert 2{\rm B},E\rangle$ 2-baryon
states with energy $E$ (and suppressed other quantum numbers).
Since the operator anomalous dimensions are mass-independent in perturbation
theory, all our results are manifestly SU$(3)$ symmetric.
The tensor product of two octets can be decomposed as
\begin{equation}
{\bf 8}\otimes{\bf 8}=({\bf 1}\oplus{\bf 8}\oplus{\bf 27})_s
+({\bf 8}\oplus{\bf 10}\oplus\overline{{\bf 10}})_a\,,
\label{six}
\end{equation}
where the first three representations are symmetric, the last three 
antisymmetric in the two baryon indices. 
In terms of octet fields, for example, we have
\begin{eqnarray}
BB^{(27)} (I=0,I_3=0,Y=0)&=& \sqrt{\frac{27}{40}}\Lambda\Lambda -\sqrt{\frac{1}{40}}\Sigma\Sigma
+\sqrt{\frac{12}{40}}N\Xi \,,\\
BB^{(8s)} (0,0,0)&=& -\sqrt{\frac{1}{5}}\Lambda\Lambda -\sqrt{\frac{3}{5}}\Sigma\Sigma
+\sqrt{\frac{1}{5}}N\Xi \,,\\
BB^{(1)} (0,0,0)&=& -\sqrt{\frac{1}{8}}\Lambda\Lambda +\sqrt{\frac{3}{8}}\Sigma\Sigma
+\sqrt{\frac{4}{8}}N\Xi \,,
\end{eqnarray}
with
\begin{eqnarray}
\Sigma\Sigma &=&  \sqrt{\frac{1}{3}}\Sigma^+\Sigma^- -\sqrt{\frac{1}{3}}\Sigma^0\Sigma^0
+\sqrt{\frac{1}{3}}\Sigma^-\Sigma^+ \\
N\Xi &=&\sqrt{\frac{1}{4}}p \Xi^- +  \sqrt{\frac{1}{4}}\Xi^- p -\sqrt{\frac{1}{4}}n \Xi^0 -  \sqrt{\frac{1}{4}}\Xi^0 n \, 
\end{eqnarray}
and
\begin{eqnarray}
BB^{(\overline{10})} (0,0,2)&=& \sqrt{\frac{1}{2}}pn -\sqrt{\frac{1}{2}}np \,,\\
BB^{(10)} (3/2,3/2,1)&=& \sqrt{\frac{1}{2}}p\Sigma^+ -\sqrt{\frac{1}{2}}\Sigma^+p\,,\\
BB^{(8a)} (0,0,0)&=& \sqrt{\frac{1}{4}}p\Xi^- -\sqrt{\frac{1}{4}}\Xi^- p
-\sqrt{\frac{1}{4}}n\Xi^0 +\sqrt{\frac{1}{4}}\Xi^0 n \,.
\end{eqnarray}
Here $I$ and $I_3$ are the total isospin and its third component while $Y$ is the total hyper charge for two baryons.
The spin quantum numbers are understood as in (\ref{physical}).
There are six s-wave potentials 
corresponding to these six channels, since, due to the total antisymmetry of
the two-baryon product, the flavor representation and
its symmetry properties uniquely determine the spin representation for
s-waves: with the antisymmetric representations the symmetric $J=1$ 
spin representation is coupled and $J=0$ is paired with the symmetric
flavor representations. In this scheme the nucleon potentials are in the
${\bf 27}$ and $\overline{\bf 10}$ channels, for isospin-spin $IJ=10$ 
and $IJ=01$, respectively.

\section{3-flavor results}

We have repeated the analysis of \cite{abw3} for the case of three quark
flavors.  In the three-flavor case some channels may become attractive 
at short distance since the Pauli exclusion principle is less significant 
than in the 2-flavor case. In this paper we therefore investigate which 
representations of the flavor SU(3) become attractive 
at short distance using the OPE.
The method and the formulae necessary to compute the 1-loop anomalous 
dimension of any local gauge invariant 6-quark operator are given in detail
in ref. \cite{abw3} and will not be reproduced here.

The local gauge invariant 6-quark operators we need are local
products of six quark field operators $q^{fa}_\alpha$ with
flavor index $f\in\{u,d,s\}$, color index $a$ and Dirac index
$\alpha\in\{1,2,3,4\}$ and can be written as local products 
\begin{equation}
T^{fgh}_{\alpha\beta\gamma}\,T^{f'g'h'}_{\alpha'\beta'\gamma'}
\label{triplets}
\end{equation}
of two color singlet triple-quark objects of the form
\begin{equation}
T^{fgh}_{\alpha\beta\gamma}=\epsilon^{abc}\,
q^{fa}_\alpha q^{gb}_\beta q^{hc}_\gamma\equiv f_\alpha g_\beta h_\gamma\,
\end{equation}
where $f,g,h,f',g',h'$ are flavor indices and $\alpha,\beta,\gamma,
\alpha',\beta',\gamma'$ Dirac indices. 
The Dirac matrices are chosen such that $\alpha=1,2$ correspond
to positive chirality and $\alpha=3,4$ to negative chirality.
As shown in \cite{abw3}, the 1-loop renormalization matrix
mixes operators with one pair of flavor and/or Dirac indices interchanged,
provided that the Dirac indices belong to the same chirality. From this
it follows that operators with different flavor or Dirac index structure do 
not mix. It also follows that (for any flavor structure) we always
have operators with vanishing RG power since for the Dirac structure 
(the square brackets indicate antisymmetrization)
\begin{equation}
T^{fgh}_{1[12]}\,T^{f'g'h'}_{3[34]}
\end{equation}
index exchange between quark fields belonging to different
triplets is not allowed by the above mentioned chirality rule and
therefore the operator has anomalous dimension 24 (coming from exchange
within the triplets), which is exactly compensated by the contribution of the
two baryons as in (\ref{betak}). 

It is sufficient to consider 6-quark operators with flavor structure
$uuddss$ since all six channels of (\ref{six}) are present
in this case. There are many possible Dirac structures but there are only
a few among these that contain \lq\lq attractive'' operators (i.e. operators
with anomalous dimension $>24$ in our units). We have considered all Dirac
structures, calculated the 1-loop anomalous dimension matrix and performed 
the change of basis necessary to make operators renormalize multiplicatively 
(diagonally) at 1-loop level. We refrain from giving a complete list here
(which would be quite long: with Dirac structure 112334 for example there are
123 independent $uuddss$ flavor 6-quark operators), and give here the complete
list of attractive operators only. These are

\begin{itemize}

\item
Dirac structure: 112334,\ \ anomalous dimension: 42\ (SU$(3)$ singlet)

\item
Dirac structure: 111222,\ \ anomalous dimension: 36\ (SU$(3)$ singlet)

\item
Dirac structure: 112234,\ \ anomalous dimension: 
36\ (SU$(3)$ singlet and octet)
 
\item
Dirac structure: 111234,\ \ anomalous dimension: 
32\ (SU$(3)$ singlet and octet)

\item
Dirac structure: 112234,\ \ anomalous dimension: 
32\ (SU$(3)$ singlet and octet)
 
\end{itemize}
We here listed operators with chiral (1,2) indices not less in number than
antichiral (3,4) ones. There are also parity reflected operators with the 
same anomalous dimension and flavor quantum numbers. 

We still have to study the question whether these operators are already present
at tree level in the OPE since otherwise their effective anomalous dimension 
is reduced by a large number (54 in our units for every extra order in 
perturbation theory with three flavors) and they become repulsive.

The baryon field operators corresponding to the flavor structure $fgh$
are built from the building blocks
\begin{equation}
b^{fgh}_{\alpha L}=f_\alpha(g_1h_2-g_2h_1)\,,\qquad
b^{fgh}_{\alpha R}=f_\alpha(g_3h_4-g_4h_3)\,,
\label{bblocks}
\end{equation}
for $\alpha=1,2,3,4$ and the physical baryon field is
\begin{equation}
B^{fgh}_\alpha=b^{fgh}_{\alpha L}+b^{fgh}_{\alpha R}+
b^{fgh}_{\hat\alpha L}+
b^{fgh}_{\hat\alpha R},
\label{physical}
\end{equation}
where $\alpha=1,2$ and  $\hat1=3$, $\hat2=4$. 
We sometimes use octet indices ($p=uud$, $\Xi^-=ssd$, etc.)
instead of the flavor notation $fgh$.

It turns out that (with the exception of the anomalous dimension 32 flavor
singlet operators) all the attractive operators are indeed present at tree 
level in the OPE for two building blocks of the form (\ref{bblocks}).
We find
\begin{eqnarray}
b^i_{\alpha L}\,b^j_{\hat\beta R}+
b^i_{\alpha R}\,b^j_{\hat\beta L} &\approx&
\delta^{ij}\,\gamma^{[42;1]}\,K^{[42]}_{\alpha\beta}+\dots ,\\
b^i_{\alpha L}\,b^j_{\beta R}+
b^i_{\alpha R}\,b^j_{\beta L} &\approx& 
f^{ijk}\,\beta^{[32;8]}\,L^{[32]k}_{\alpha\beta}+
\delta^{ij}\,\beta^{[36;1]}\,\epsilon_{\alpha\beta}\,L^{[36]}\nonumber\\
&\qquad\qquad +& d^{ijk}\,\beta^{[36;8]}\,\epsilon_{\alpha\beta}\,L^{[36]k}
+\dots,\\
b^i_{\alpha L}\,b^j_{\beta L} &\approx&
\delta^{ij}\,\alpha_1^{[36;1]}\,\epsilon_{\alpha\beta}\,M^{[36]}+\dots,\\
b^i_{\hat\alpha L}\,b^j_{\hat\beta L} &\approx&
\delta^{ij}\,\alpha_2^{[36;1]}\,\epsilon_{\alpha\beta}\,L^{[36]}+
d^{ijk}\,\alpha^{[36;8]}\,\epsilon_{\alpha\beta}\,L^{[36]k}
+\dots,
\end{eqnarray}
where $\dots$ stand for non-attractive operators. We also have the parity
reflected counterparts of the above relations. We will denote the parity
reflected operators with a $\hat{\phantom{L}}$. Here the coefficients
$\gamma, \beta, \alpha$ are nonvanishing and the $K,L,M$ are local 6-quark
operators with their anomalous dimension and SU$(3)$ quantum numbers indicated.

Finally we can compute the baryon-baryon tree level OPE:
\begin{equation}
B^i_\alpha B^j_\beta \approx \delta^{ij}\,\epsilon_{\alpha\beta}\,
\left\{ a_1{\cal A}_1^{[36]}+a_2{\cal A}_2^{[36]}+a_3{\cal A}_3^{[42]}\right\}
+d^{ijk}\,\epsilon_{\alpha\beta}\,b\,{\cal B}^{[36]k}+f^{ijk}\,c\,
{\cal C}_{\alpha\beta}^{[32]k}+\dots
\end{equation}
Here
\begin{equation}
\begin{split}
a_1&=\alpha_1^{[36;1]},\\
a_2&=\alpha_2^{[36;1]}+\beta^{[36;1]},\\
a_3&=\gamma^{[42;1]},\\
b&=\alpha^{[36;8]}+\beta^{[36;8]},\\
c&=\beta^{[32;8]},
\end{split}
\qquad\quad
\begin{split}
{\cal A}_1^{[36]}&=M^{[36]}+\hat M^{[36]},\\
{\cal A}_2^{[36]}&=L^{[36]}+\hat L^{[36]},\\
{\cal A}_3^{[42]}&=K^{[42]}_{12}-K^{[42]}_{21},\\
{\cal B}^{[36]k}&=L^{[36]k}+\hat L^{[36]k},\\
{\cal C}^{[32]k}_{\alpha\beta}&=L^{[32]k}_{\alpha\beta}
+\hat L^{[32]k}_{\alpha\beta}.
\end{split}
\end{equation}

We see that there are no attractive operators in the ${\bf 27}_s$, ${\bf 10}_a$
and $\overline{\bf 10}_a$ channels. This is consistent with what we found in 
the nucleon case; these belong to ${\bf 27}_s$ and $\overline{\bf 10}_a$.
There are three attractive channels: ${\bf 1}_s$, ${\bf 8}_s$, ${\bf 8}_a$.
Among these the NBS wave function for ${\bf 1}_s$ becomes most singular at short distance since its anomalous dimension 42 is largest. In terms of the corresponding potential, however, all three have the same $r$ dependence at short distance.
They may not be equally strong: the coefficients $a_1$ etc. can be 
calculated\footnote{e.g. at tree level $\alpha^{[36,8]}/\beta^{[36,8]}=-2/3$ 
and $\alpha_2^{[36,1]}/\beta^{[36,1]}=1/3$.}
but these coefficients are to be multiplied by the matrix elements of the
corresponding operators, which we do not know.

We note that there is no local 6-quark operator in the ${\bf 8}_s$ channel
with nonvanishing non-relativistic limit, and therefore the corresponding
matrix elements also vanish in the same limit.
We considered a class of 6-quark operators in the non-relativistic
limit, which have $uuddss$ flavor and $S_z=0$ with $S_z$ being the $z$
component of the total spin.
Since $q_\alpha = q_{\hat\alpha}$ in the non-relativistic limit, this class
of operators has $111222$ Dirac structure.
We found by an explicit calculation that all attractive 6-quark operators
in the ${\bf 8}_s$ channel vanish in the non-relativistic limit.
Next we calculated the
eigenvalues $T$ of the SU(3) Casimir operator (and their eigenvectors) for
this class of operators. The octet representation corresponds to $T=3$.
We have found that while the operator $p_1\Xi^-_2-\Xi^-_1p_2$ contains $T=3$
components,
the operator $p_1\Xi^-_2+\Xi^-_1p_2$ does not contain $T=3$ components at
all. (The subscripts $1,2$ represent spin indices.)
This shows that there is no ${\bf 8}_s$ 6-quark operator in the non-relativistic limit.

One can also see the above property  group theoretically\cite{OSY}.
The 6-quark operator must be totally antisymmetric  due to the Fermi-Dirac statistics.
In terms of the permutation group, the totally antisymmetric representation corresponds to $[1^6]$, whose Young diagram has 6 boxes in a column.
Each quark operator in the non-relativistic limit has color($c$), spin($s$), flavor($f$) in addition to the coordinate $x$ which is common for all 6 quarks in the local operator. We now consider symmetries of each index. The symmetry for color SU(3) must be $[2^3]_c$ due to gauge invariance, while
that for the coordinate is totally symmetric, therefore $[6]_x$. For 3 flavor, the non-relativistic quark belongs to the fundamental representation  of SU(6)$_{sf}\supset$ SU(2)$_s \otimes$ SU(3)$_f$. The symmetry of 6 quarks in SU(6)$_{sf}$, denoted as $[F]_{sf}$, must satisfy
\begin{eqnarray}
[2^3]_c \otimes [F]_{sf} \otimes [6]_x \supset [1^6] \Rightarrow [2^3]_c \otimes [F]_{sf}  \supset [1^6] ,
\label{eq:condA}
\end{eqnarray}
 in order to make the 6-quark operator totally antisymmetric.
 In SU(6)$_{sf}$,  $[3]$ contains octet and decuplet, and therefore 
 two octet baryon state belongs to $[3]\otimes [3] = [6]\oplus [4,2] \oplus [5,1]\oplus [3^2] $, where the first(last) two are (anti)symmetric under exchange of two $[3]$. 
 Since two baryons at the same space-time point must be antisymmetric under exchange of the two, $[5,1]$ and $[3^2]$ are allowed. 
The table C on p.116 of Ref.\cite{IN} tells us
\begin{eqnarray}
 [2^3]_c \otimes [5,1]_{sf}  = [2^2, 1^2] \oplus  [3,2,1] ,\quad
 [2^3]_c \otimes [3^2]_{sf}  = [1^6] \oplus  [2^2, 1^2] \oplus  [3^2]\oplus  [4, 1^2]  .
\end{eqnarray}
These show that  the $[5,1]_{sf}$ representation does not satisfy the condition (\ref{eq:condA}) while $[3^2]_{sf}$ does. Therefore  only $[3^2]_{sf}$, but not $[5,1]_{sf}$, must appear in the local (gauge-invariant) 6-quark operator. In terms of SU(6)$_{sf}$, the 6 quarks in the ${\bf 8}_s$ channel, which has $[3^2]_s$ in spin and $[3,2,1]_f$ in flavor, becomes (see also the same table C)
\begin{eqnarray}
{\bf 8}_s = [3^2]_s\otimes [3,2,1]_f = [5,1] \oplus [4,2] \oplus [4,1^2]\oplus 2 [3,2,1] \oplus 3 [3,1^3]\oplus [2^2,1^2] \oplus [2,1^4]  \nonumber\\  
\end{eqnarray}
while ${\bf 1}_s$ and ${\bf 8}_a$ are expressed as
\begin{eqnarray}
{\bf 1}_s =  [3^2]_s\otimes [2^3]_f &=& [3^2] \oplus [4,1^2]\oplus [2^2,1^2] \oplus [1^6]
\end{eqnarray}
\begin{eqnarray}
{\bf 8}_a=[4,2]_s\otimes [3,2,1]_f  
&=& [5,1] \oplus 2 [4,2] \oplus 2 [4,1^2]\oplus [3^2] \oplus 3 [3,2,1] \nonumber \\
&& \oplus   [2^3]\oplus 2[3,1^3] \oplus 2[2^2,1^2]   \oplus [2,1^4]  .  
\end{eqnarray}
Since ${\bf 8}_s$ contains $[5,1]$ only but no $[3^2]$, the 6 quark operator in this channel does not exist in the SU(6)$_{sf}$ representation (i.e. the non-relativistic limit). 

\section{Comparison to quark model calculations and lattice Monte Carlo}

The short distance part of the baryon potential has also been calculated
using a valence quark model with gluon exchange \cite{OSY} and lattice
Monte Carlo measurements \cite{su3MC}.
 
In the quark model above the only attractive channel is the SU$(3)$ singlet one
(which may correspond to a strange dibaryon bound state: the $H$ particle).
It is not possible to quantitatively compare the results with ours since the 
way the baryon potential is defined in the two formalisms is different and 
in the valence quark model SU$(3)$ breaking effects are also taken into 
account (and they are, in some cases, comparable to the SU$(3)$ symmetric 
contributions). In addition, a particle basis ($N,\Xi,\Sigma,\Lambda$) is 
used in the valence quark model and the various s-wave channels are labeled 
as in this example:
\begin{equation}
N\Sigma\qquad\quad (I,J)=(\frac{1}{2},0)\,.
\end{equation}
In the quark model this channel is for example strongly repulsive but this 
result cannot be directly compared to ours since in our language this is a 
linear combination of the channels ${\bf 27}_s$ and ${\bf 8}_s$. On the
other hand, for the case
\begin{equation}
N\Sigma\qquad\quad (I,J)=(\frac{3}{2},0)
\end{equation}
the quark model gives (a somewhat weaker) repulsion, consistently with our 
result, since this channel is pure ${\bf 27}_s$.

\begin{figure}
\begin{center}
\psfig{figure=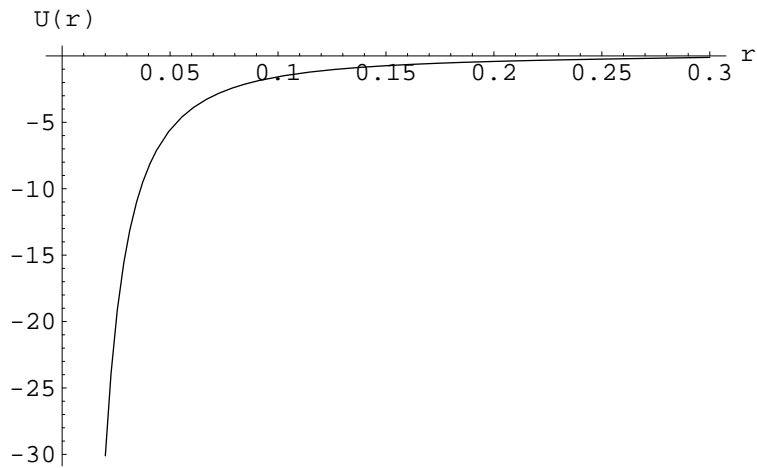,width=10cm}
\end{center}
\vspace{-0.5cm}
\caption{\footnotesize 
Potential corresponding to the model wave function with
$\nu_1=2/9$, $\nu_2=-0.10$ and $R=2$.
}
\label{potpc2}
\end{figure}

The MC measurements \cite{su3MC} are performed using full QCD with three
massive quarks in the flavor SU$(3)$ limit. The results (for s-wave potentials)
are analysed in terms of the six channels (\ref{six}) and attraction is
found only in the singlet channel. The other five are found to be repulsive,
with four of them (including the ${\bf 27}$ and $\overline{\bf 10}$ already
occurring in the nucleon case) being rather similar. The potential 
corresponding to the ${\bf 8}_s$ channel is different from all the other ones
in that it does not have any attractive pocket and its repulsive core is 
stronger and broader than for the other cases.  

Our results do not necessarily contradict to what was found in the quark model
and measured on the lattice since the distance range in which our asymptotic
results are valid may not overlap with the range where the other two methods
work. The latter is the range between $0.1$ fm and $1.5$ fm, whereas the true
asymptotic behavior predicted by asymptotically free perturbation theory may
be valid in a much shorter distance range only and in the range where 
comparison is possible subleading contributions in the OPE may be 
non-negligible. To illustrate this situation we consider a (renormalization
group motivated) simpified model where the baryon wave function is the sum 
of a leading term with RG power $\nu_1$ and a subleading term with relative
coefficient $R$ and RG power $\nu_2$. More precisely, we take
\begin{equation}
\psi(r)=\lambda^{-\nu_1}+R\,\lambda^{-\nu_2}
\label{model}
\end{equation}
as our simplified baryon wave function, where $\lambda(r)$ is the 2-loop
3-flavor running coupling defined by
\begin{equation}
\frac{1}{\lambda}+\kappa\ln\lambda=\ln\frac{r_0}{r}\,,\qquad
\kappa=\frac{\beta_1}{2\beta_0^2}=\frac{32}{81}.
\end{equation}
Using $\lambda(r)$ instead of $1/\ln(r_0/r)$ takes into account some of the
higher perturbative contributions. Of course, the complete RG asymptotic form
of the NBS wave function is much more involved than the simple formula 
(\ref{model}). There are perturbative corrections (power series in $\lambda$)
multiplying both the leading power of $\lambda$ and the subleading one and
there are also many sub-subleading terms to be added. Nevertheless the simple
model (\ref{model}) can qualitatively reproduce the main features of the true
asymptotics of the wave function. We take
\begin{equation}
\nu_1=\frac{2}{9}\,,\qquad\quad \nu_2=-0.1\,.
\end{equation}
We have chosen these numbers because we want to model the ${\bf 8}_s$ 
channel, which is asymptotically attractive but looks strongly repulsive 
in the range studied by lattice Monte Carlo and valence quark models.
$\nu_1$ corresponds to anomalous dimension 36, which is the largest 
(and only attractive one) occurring in this channel and $\nu_2$ is chosen
to represent the \lq\lq average'' of the RG power zero (which, as we have seen,
occurs in all channels) and the many negative sub-subleading powers.

\begin{figure}
\begin{center}
\psfig{figure=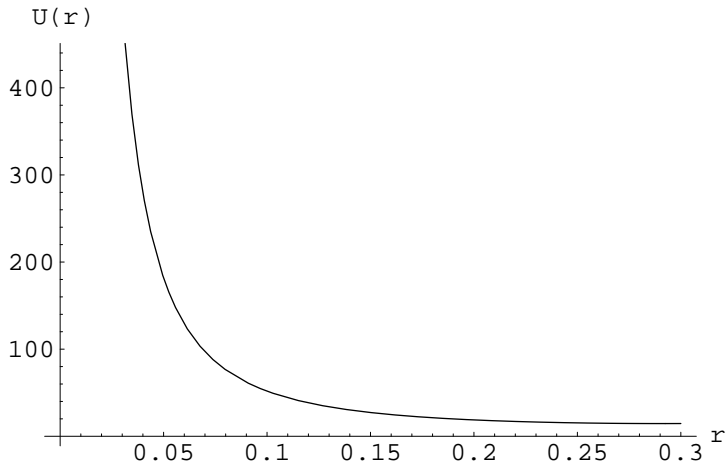,width=10cm}
\end{center}
\vspace{-0.5cm}
\caption{\footnotesize 
Potential corresponding to the model wave function with
$\nu_1=2/9$, $\nu_2=-0.10$ and $R=-2$.
}
\label{potpcm2}
\end{figure}

Clearly for $r\to0$ eventually the leading term dominates, but
since $\nu_1$ and $\nu_2$ are both small numbers the subleading effects
are still important in the range 0.1-0.3 fm, where the short distance expansion
(\ref{model}) can be compared to results obtained by the above two methods.
If the relative coefficient $R$ becomes large, the second term will 
dominate in this range and the potential becomes repulsive. 
In Fig. \ref{potpc2} the potential\footnote{The model potential is simply
defined as $U(r)=\nabla^2\psi/\psi$. We are using $r_0$ units for the plots.} 
for $R=2$ is shown. It is less attractive
than  the one corresponding to the leading term alone, but is still weakly
attractive. On the other hand for $R=-2$ already the second term dominates and
the potential, which is shown in Fig. \ref{potpcm2}, is clearly repulsive in
this range.

A natural question arising is that if this mechanism explains why an 
asymptotically attractive potential can look repulsive in the intermediate
range 0.1-0.3 fm, why the same mechanism can not modify the results 
obtained in the nucleon case (where repulsion was found in agreement with
the quark model and Monte Carlo results). The answer is that in the nucleon
case all subleading terms correspond to repulsion so if they happen to dominate
in the intermediate range the qualitative behavior of the potential remains
unchanged. Actually the nucleon case corresponds to the degenerate case with
$\nu_1=0$ and the true asymptotics is determined by the sign of the relative
coefficient $R$. But the above explained mechanism actually improves the
situation in the intermediate range since if $R$ becomes large (with either
sign), it dominates and the potential looks repulsive.
Note that the potential  should become attractive asymptotically at short distance for negative $R$.
The attractive potential can be seen in the intermediate range only if $R\approx-1$ or smaller.

In addition to the above mechanism, the matrix elements of the local 6 quark operator
in the ${\bf 8}_s$ channel vanish in the non-relativistic limit, as shown in the previous section.
Therefore the attraction in the ${\bf 8}_s$ channel predicted by the OPE analysis can be hardly seen, as long as the non-relativistic approximation holds. Indeed this approximation seems good
in the MC measurements\cite{su3MC} since the wave function in the ${\bf 8}_s$ channel is observed to be very small at the origin.  

\section{Conclusion}
In this paper we extended the OPE analysis of ref. \cite{abw3} from two 
to three flavors, in order to investigate the short distance behavior of 
general baryon-baryon potentials in flavor SU(3). 
Interestingly the OPE analysis leads to manifest attraction for the SU(3) 
singlet potential at short distance, which is consistent with the quark model 
prediction and the lattice QCD result.
The manifest attraction channel does not exist in the nucleon-nucleon 
potential for 2 flavors (the channels ${\bf 27}_s$ and $\overline{\bf 10}_a$ 
in the 3 flavor theory). The increase of the number of flavors decreases the 
repulsion at short distance and in some cases turns it into attraction 
already for 3 flavors, since the Pauli exclusion principle among quarks 
becomes less significant for larger number of flavors.  The agreement among 
the three different methods strongly corroborates that the interaction between 
two baryons in the singlet channel is indeed attractive.
It would be important to check by a lattice QCD calculation 
whether this attraction in the flavor SU(3) limit leads to a bound-state 
(the H-particle). Even if the bound state is indeed present in the SU$(3)$
limit, it is possible that it becomes a resonance above the $\Lambda\Lambda$ 
threshold in the real world where the strange quark is heavier than 
the other two.

On the other hand, our OPE analysis indicates that there is short distance 
attraction in both (${\bf 8}_a$ and ${\bf 8}_s$) octet channels, which 
disagrees with present lattice QCD results.  
Although we gave a possible explanation 
for this discrepancy, a direct confirmation by lattice QCD calculations will 
be needed for definite conclusions.  

\section*{Acknowledgments}
S.A. would like to thank members of HAL QCD Collaboration, in particular, Prof. T. Hatsuda, Drs. Doi, Inoue, Ishii, Nemura and Sasaki, for useful discussions. 
S. A. is supported in part by Grant-in-Aid of the Ministry of Education, 
Sciences and Technology, 
Sports and Culture (Nos. 20340047, 20105001, 20105003).
J. B. is grateful to the Center for Computational Sciences of
the Tsukuba University for financial support and to the Particle Physics
group for hospitality during his stay.  
This investigation was supported in part by the Hungarian National 
Science Fund OTKA (under K77400).


\end{document}